\documentclass[10pt]{article}
\usepackage[left=2cm, right=2cm, top=3cm]{geometry}
\usepackage{tocloft,hyperref,graphicx,multirow,parskip,ragged2e,amsmath,makecell}
\usepackage{fancyvrb,array, multirow,watermark,tablefootnote}
\hypersetup{
    colorlinks=true,
    linkcolor=blue,
    filecolor=magenta,      
    urlcolor=blue,
}
\makeatletter
\newcommand*{\rom}[1]{\expandafter\@slowromancap\romannumeral #1@}
\makeatother
\title{\vspace{-3em}{\Huge {\fontfamily{ptm}\selectfont Fearless Steps Challenge (FS-\rom{1})} \vspace{0.5em}\\\LARGE {\fontfamily{ptm}\selectfont 2019 Evaluation Plan}\vspace{-0.5em}}}
\author{Aditya Joglekar, John H.L. Hansen}

\author{\vspace{1em}{\large {\fontfamily{ptm}\selectfont Aditya Joglekar, John H.L. Hansen}}}

\date{\vspace{-3ex}}
\usepackage[titletoc,title]{appendix}
\usepackage{enumitem,courier}

\usepackage[table]{xcolor} 
\usepackage[bottom]{footmisc}

\usepackage{fancyhdr}
\usepackage{lastpage}
\definecolor{lightgray}{gray}{0.9}
\newcolumntype{C}[1]{>{\centering\arraybackslash}m{#1}}
\newcolumntype{L}[1]{>{\flushleft\arraybackslash}m{#1}}
\newcolumntype{R}[1]{>{\flushright\arraybackslash}m{#1}}
\begin{document}

\maketitle \vspace{-1em}
\renewcommand*\contentsname{}

\begin{figure}[!ht]
\Centering
\includegraphics[width=12.5cm, height=16cm]{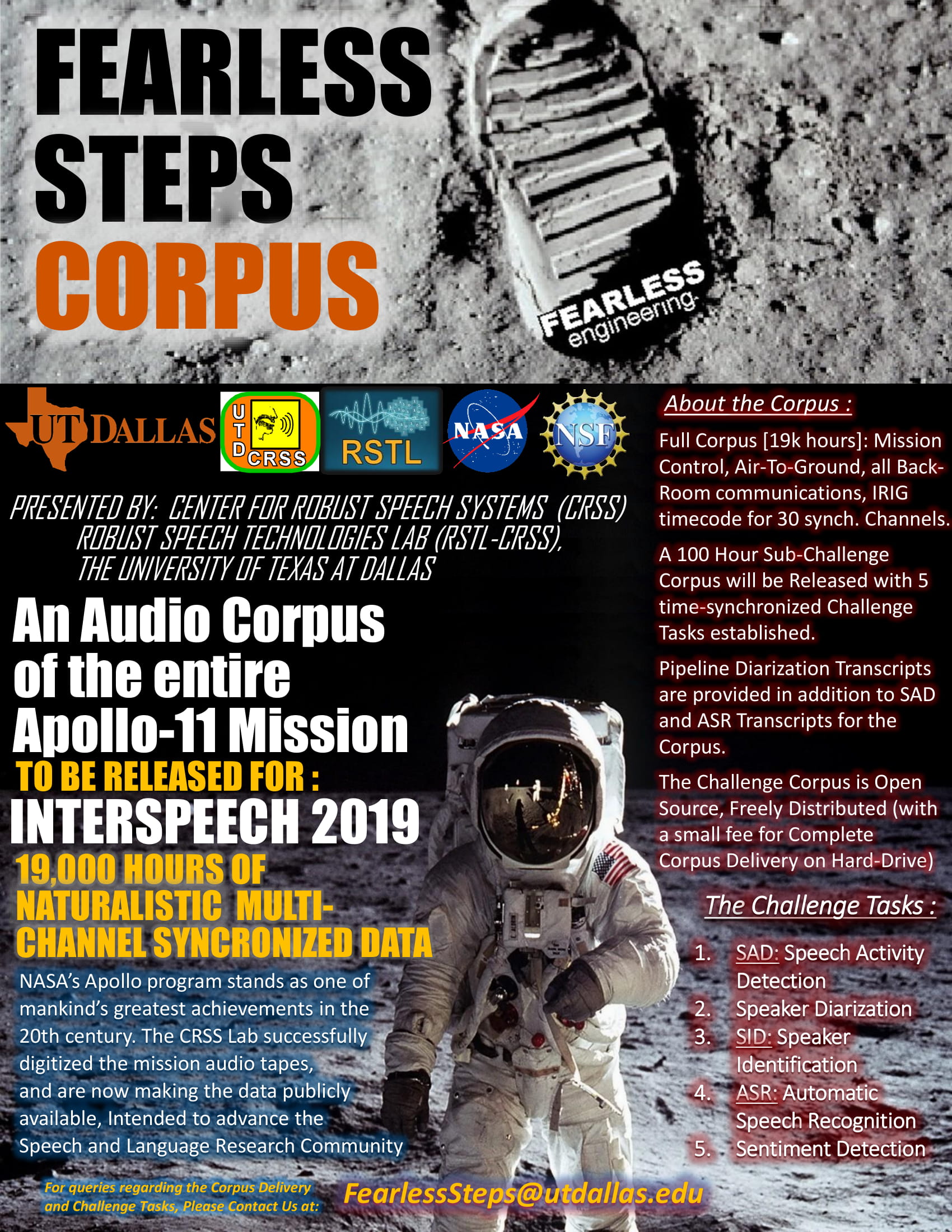}
\end{figure}
\newpage
\date{Last Edited: May 20, 2019}
\tableofcontents  
\thispagestyle{fancy}

\vspace{35em} 
\fontsize{12}{8}\selectfont
\textit{Disclaimer: The Fearless Steps corpus is derived from a five-year NSF CISE funded project awarded to CRSS at the University of Texas at Dallas. UTDallas-CRSS established the hardware/software solutions to digitize and diarize 19,000 hrs of NASA Apollo data. All core Apollo data released as part of this challenge has been approved for public release by NASA Export Control. The full audio corpus is also available from UTDallas-CRSS. Any mention of or reference to organizations other than UTD is for information only; it does not imply recommendation or endorsement by UTDallas-CRSS nor does it imply that the products mentioned are necessarily the best available for the purpose. }
\newpage
\fontsize{12}{14}\selectfont

\section{Introduction}

The Fearless Steps Challenge 2019 Phase-1 (FS-1 2019) is the inaugural Challenge of the Fearless Steps Initiative hosted by \href{https://crss.utdallas.edu/}{Center for Robust Speech Systems (CRSS)} at the University of Texas at Dallas. The goal of this Challenge is to evaluate the performance of state-of-the-art speech and language systems for large task oriented teams with naturalistic audio in challenging environments. 

Researchers may select to participate in any single or multiple of these challenge tasks. Researchers may also choose to employ the FEARLESS STEPS corpus for other related speech applications. All participants are encouraged to submit their solutions and results for consideration to this ISCA INTERSPEECH-2019 special session.

While there is an extensive amount of audio (19,000 hrs), the core data for this FEARLESS STEPS challenge is drawn from 5 channels, over three Apollo-11 phases consisting of 100 hours. (see Section-\ref{sec:Data} for more details). A total of 80 hours of audio are provided for task system development. For these 80 hours, a sub-set of 20 hours of human verified ground truth labels and transcripts are provided (SAD, SD, SID, ASR). An additional set of 20 hours will be released for open test evaluation (see Section-\ref{sec:Timeline}).

This (FS-1) edition of the FEARLESS STEPS Challenge includes the following five tasks:

\begin{description}[style=multiline,leftmargin=4cm,font=\normalfont]
    \item[\hspace{0.5cm}$\bullet$ \textbf{Task \#1}] Speech Activity Detection        (\textbf{SAD})
    \item[\hspace{0.5cm}$\bullet$ \textbf{Task \#2}] Speaker Diarization              (\textbf{SD})
    \item[\hspace{0.5cm}$\bullet$ \textbf{Task \#3}] Speaker Identification           (\textbf{SID})
    \item[\hspace{0.5cm}$\bullet$ \textbf{Task \#4}] Automatic Speech Recognition     (\textbf{ASR})
    \item[\hspace{0.5cm}$\bullet$ \textbf{Task \#5}] Sentiment Detection              (\textbf{SENTIMENT})
\end{description}

\section{Objectives}
Traditionally, most speech and language technology concentrates on analysis of a single audio stream or channel with one or more speakers involved. The Apollo audio data represents 30 individual analog communications channels with multiple speakers in different locations working real-time to accomplish NASA's Apollo missions. For Apollo-11, this means each channel reflects a single communications loop (channel) that can contain anywhere from 3-33 speakers over extended time periods. While each channel has a primary function with a specific NASA Mission Specialist responsible, each of these channels are ``loops", which contain core speakers working together plus speech from background conversations looped in at times, some reflecting Air-to-Ground (CAPCOM - Capsule Communicator) communications from the Astronauts. In addition, vast majority of the original Apollo Mission analog audio are all unlabeled, making application of speech technology a challenge. The inaugural phase of the Challenge (FS-1) will therefore emphasize the need to address various single channel speech tasks using unsupervised and/or semi-supervised speech algorithms. The Challenge Tasks for this session encourage the development of such solutions for core speech and language tasks on data with limited ground-truth/low resource availability, and serves as the ``First Step'' towards extracting high level information from such a task driven unlabeled corpus. \par
\vspace{-1em}
\section{Tentative Schedule}
\begin{center}
    \rowcolors{2}{lightgray}{white}
    {\renewcommand{\arraystretch}{2}%
    \begin{tabular}{L{8cm}R{7cm}}
    \rowcolor{gray!20}
    \large Registration Period & \large \textbf{February 11 - April 6, 2019} \\
    \large Evaluation Plan Release & \textbf{\large  February 11, 2019} \\
    \large Training and Development set release & \textbf{\large February 11, 2019} \\
    \large Baseline Results (Dev set) & \textbf{\large February 16, 2019} \\
    \large Evaluation set release & \textbf{\large March 1, 2019} \\
    \large Baseline Description and Results (Eval set) & \textbf{\large March 20, 2019} \\
    \large System Submission Opens & \textbf{\large March 14, 2019} \\
    \large Interspeech Paper Registration deadline & \textbf{\large March 29, 2019} \\
    \large Interspeech Paper Submission deadline & \textbf{\large April 5, 2019} \\
    \large Final System Submission Deadline & \textbf{\large June 28, 2019} \\
    \large Final Results Announced for all Tasks & \textbf{\large July 20, 2019 at 9:56pm CST}  \\[-0.5ex]
    \rowcolor{white} \multicolumn{2}{c}{ \Gape[0.10cm][0.10cm]{\color{red} \textbf{\textit{\large (The exact time of the First Moon Walk, 50$^{th}$ Anniversary)}}}}\\
    \rowcolor{lightgray}\large Interspeech 2019 Special Session & \textbf{\large September 15 - 19, 2019} \\
    \end{tabular}}
\end{center}



\section{Data Overview}
\label{sec:Data}
\subsection{Corpus Development}
\begin{figure}[!ht]
\begin{center}
  \includegraphics[width=14cm, height=6cm]{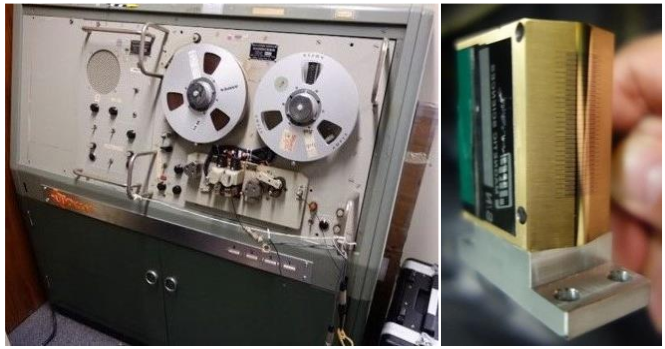}
  \caption{(left): The SoundScriber device used to decode analog tapes, and (right): The UTD-CRSS designed read-head decoder retrofitted to the SoundScriber}
  \label{fig:Soundscr}
\end{center}
\end{figure}

The Apollo 11 mission lasted 8 days 3 hours 18 minutes and 35  seconds. The entire communications between astronauts, flight controllers,  and  their  backroom  support  teams  inside  NASA Mission  Control  Center  (MCC)  were  continuously  recorded using  two  30-track  analog  reel-to-reel  recording  machines, namely Historical Recorder 1 (HR1) and 2 (HR2). By alternately changing the tapes, continuity was ensured without any loss of data; 29 of the 30 channels/tracks on the analog tape were used to record speech data with one channel recording  the  Mission  Elapsed  Time  (MET)  in  an  encoded  IRIG-B format. The records stored by the United States National Archives and Records Administration (NARA) were used to digitize the original analog tapes, by designing a new read-head for the SoundScriber player (as shown in Fig. \ref{fig:Soundscr}). The read-head was developed specifically with the aim of digitizing all the channels simultaneously, thus preserving the synchronicity of the data, enabling individual channels from each HR1 and HR2 to be indexed and stored separately. 

Synchronous multichannel reconstruction of the entire mission is made possible by using the first channel of every tape which contains the MET. This data was stored and digitized initially at a 44.1 kHz sampling frequency, and later downsampled to 8 kHz for speech analysis. The recordings were saved as half-hour chunks per channel with their file names indicating the mission name, the historical recorder and channel the recording belongs to, followed by the start and end times as given by the mission elapsed time.

\begin{figure}[!ht]
\begin{center}
  \includegraphics[width=15cm, height=9cm]{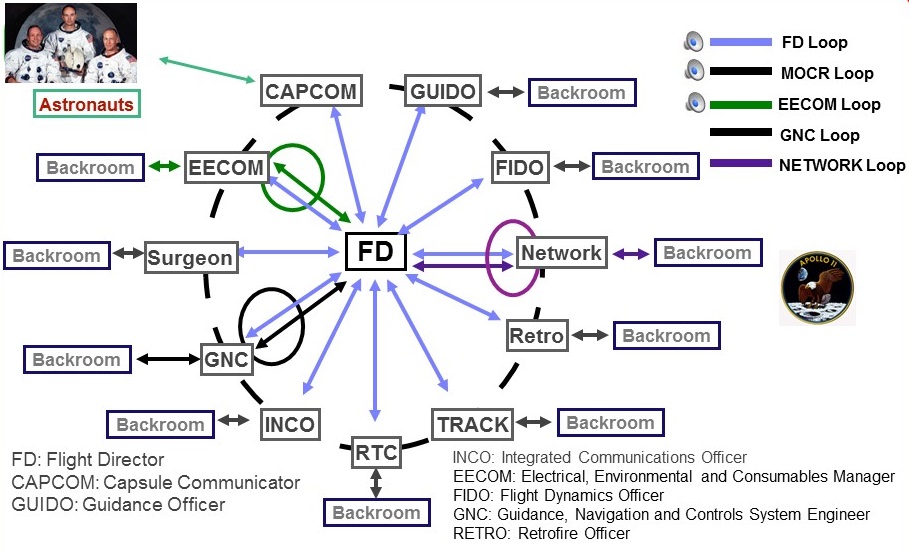}
  \caption{Apollo-11 communication overview. Ground communications between hundreds of flight controllers and their ’backroom’ support staff are shown in the loop. The space-to-ground communications are linked to CAPCOM}
  \label{fig:commProc}
\end{center}
\end{figure}

\textbf{Communication Protocols}: Specific protocols followed during the mission were imperative for ensuring successful communication. Knowledge of these communication characteristics can be leveraged to achieve better inferences through informed analysis. Fig. \ref{fig:commProc}. shows a basic structure of the communication protocols. Only the Capsule Communicator (CAPCOM) could directly communicate with the astronauts, with the Flight Director (FLIGHT) in control of accessing all other channel loops. With multiple speakers joining in on these loops at various points in time, all personnel would address the channel owner by their assigned channel names. In fact, all backroom staff are present on multiple channel loops.  Audio markers such as `Quindar Tones' were used to infer communication with the astronauts.

\subsection{Data Organization}

\begin{figure}[!ht]
\begin{center}
  \includegraphics[width=12cm, height=7cm]{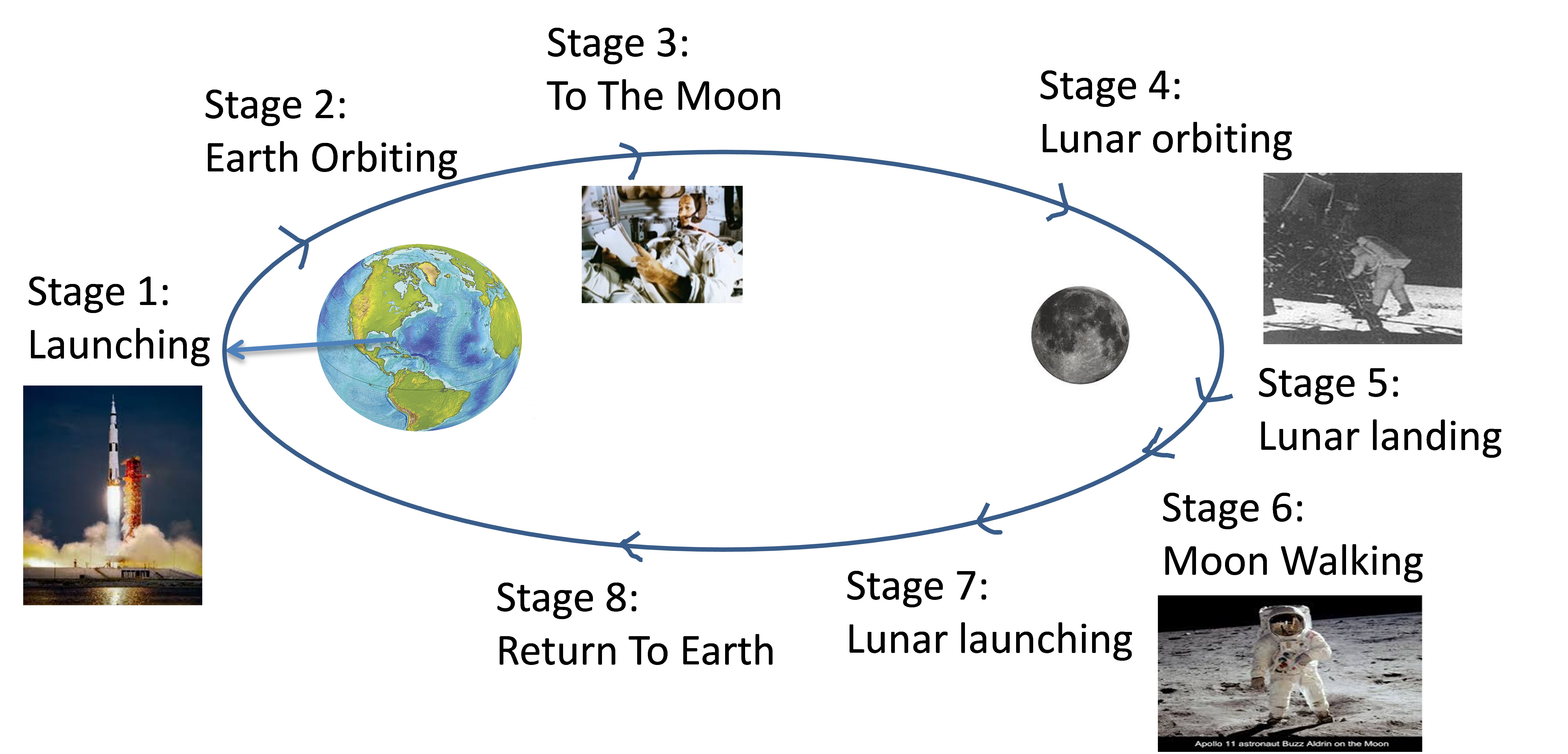}
  \caption{Overview of the timeline of Apollo 11 mission}
  \label{fig:a11timl}
\end{center}
\end{figure}

Fig. \ref{fig:a11timl}. displays the overall Timeline of the Apollo-11 Mission. The Stages ‘1’, ‘5’ and ‘6’ which were high impact mission-critical events were found to be ideal for the development of the 100-hour Challenge Corpus. With the quality of speech data varying between 0 and 20 dB SNR in this challenge corpus, the channel variability and complex interactions across all five channels of interest discussed in the previous section are mostly encapsulated in these 100 hours. \\ \\

 These three major events the multichannel data is chosen from are: 
\begin{itemize}
    \item Lift Off (25 hours)
    \item Lunar Landing (50 hours)
    \item Lunar Walking (25 hours)
\end{itemize}
\
These landmark events have been found to possess rich information from the speech and language perspective. Out of the 29 channels, five channels of interest with the most activity over the selected events were chosen to select the data from:
\begin{itemize}
    \item  Flight Director (FD)
    \item  Mission Operations Control Room (MOCR)
    \item  Guidance Navigation and Control (GNC)
    \item  Network Controller (NTWK)
    \item  Electrical, Environmental and Consumables Manager (EECOM)
\end{itemize}
The personnel operating these five channels (channel owners/primary speakers) were in command of the most critical aspects of the mission, with additional backroom staff looping in for interactions with the primary owners of this channel.

\subsection{Challenges with the Apollo Data}
\label{subsec:DataChallenges}
The Apollo data is unique and poses multiple challenges. Comprised of 19,000 hours of naturalistic multi-channel data, it is characterized by multiple classes of noise and degradation and several overlap instances over most channels. Most audio channels are degraded due to high channel noise, system noise, attenuated signal bandwidth, transmission noise, cosmic noise, analog tape static noise, noise due to tape aging, etc. Moreover, the noise conditions and signal-to-noise ratio (SNR) levels vary over a 25 dB range for separate channels, and in many cases, at different mission stages within each channel (See Tab. \ref{tab:SNRvals}). Some channels even have the presence of babble noise depending on the location of the personnel in the MCC. For the Apollo missions, head-mounted Plantronics microphones were used, but some in-spacecraft recordings were made using fixed far-field microphones which also picked up the presence of environmental noise (e.g., glycol cooling pumps and thruster firings) that varied over time. These conditions severely degrade the performance of conventional speech activity detection algorithms. Due to the time-critical nature of the mission, multiple instances with rapid switching of speakers exist. In many cases, extremely short duration responses are common. As an example, the average duration for speakers during mission status updates are close to 0.5 seconds with as many as 15 speakers occurring in turns in a span of 10 seconds. This poses a serious challenge for speaker recognition and diarization systems. The speech density varies dramatically over time, depending on the channel, the stage of the mission, and issues encountered during the mission. Some instances show more than 20 active speakers at a time, carrying conversations for 15 minutes at a stretch, and some other instances show extended periods of silence, usually for hours. For astronauts, their vocal tract characteristics have been observed to have considerable changes through different stages of the mission. These factors render thresholding mechanisms and diarization system performances degraded. The conversational content in the missions was specific to the standards maintained by NASA for efficient air-to-air, air-to-ground, and ground communications. Using standard language models for this application can lead to misclassification of words detected, posing a significant challenge for speech recognition. Hence, there is a need for incorporating NASA specific vocabulary to existing language models. Analyzing unprompted speech has its own challenges. All the speech recorded in the corpus is unprompted, and hence subject to significant variations in speech characteristics for every speaker. These are some challenges that have shown to significantly degrade the quality of generalized SAD, ASR, SID, and Diarization models. Apollo specific application developments have shown to drastically improve system accuracy on these models. \footnote{\href{https://www.isca-speech.org/archive/Interspeech_2018/pdfs/1942.pdf}{Fearless Steps Overview Paper Interspeech 2018}}

\subsection{General Statistics}

Due to the communication characteristics observed for the audio data, there is a presence of background conversation speech in most of the audio. Previous efforts have concentrated on analysis of only primary conversation speech, which is what is presented below. The distribution of total primary conversation speech content in each of the channels for every event has been given in Table \ref{tab:totSpeechDur}, with the total content for each event provided in the final column, and over each channel provided in the final row.

\begin{table}[!ht]

\centering
{\renewcommand{\arraystretch}{1.2}%
\begin{tabular}{ |c|c|c|c|c|c|c| }
\hline
& \textbf{EECOM} & \textbf{FD} & \textbf{GNC} & \textbf{MOCR} & \textbf{NTWK} & \textbf{Total}\\ \hline
\textbf{Lift Off} & 2.1 & 1.2 & 1.3 & 0.8 & 3.9 & 9.3\\ \hline
\textbf{Lunar Landing} & 3.7 & 1.3 & 4.0 & 0.9 & 4.4 & 14.3\\ \hline 
\textbf{Lunar Walking} & 3.9 & 1.1 & 3.0 & 1.4 & 2.8 & 12.2\\ \hline
\textbf{Total} & 9.7 & 3.6 & 8.3 & 3.1 & 11.1 & 35.8\\ \hline
\end{tabular}}
\caption{Total Speech Durations (hours) per Channel and Event}
\label{tab:totSpeechDur}
\end{table}


To make sure there is an equitable distribution of data into training, test, and development sets for the challenge tasks, we have categorized the data based on noise levels, amount of speech content, and amount of silence. Due to the long silence durations, and based on importance of the mission, the speech activity density of the corpus varies throughout the mission.

\begin{table}[!ht]
\centering
{\renewcommand{\arraystretch}{1.2}%
\begin{tabular}{ |c|c|c|c|c|c| }
\hline
& \textbf{EECOM} & \textbf{FD} & \textbf{ GNC } & \textbf{MOCR} & \textbf{NTWK} \\ \hline
\textbf{SNR (Mean)} & 13.32 & 14.67 & 14.91 & 5.07 & 10.68\\ \hline
\textbf{SNR (Std. Dev)} & 7.40 & 10.51 & 11.96 & 12.60 & 11.17\\ \hline 
\end{tabular}
\caption{Signal to Noise Ratio Statistics (dB SNR) per channel for Dev Data}
\label{tab:SNRvals}}
\end{table}

\begin{table}[!ht]
\centering
{\renewcommand{\arraystretch}{1.2}%
\begin{tabular}{ |c|c|c|c|c|c| }
\hline
& \textbf{EECOM} & \textbf{FD} & \textbf{GNC} & \textbf{MOCR} & \textbf{NTWK}\\ \hline
\textbf{Avg. Num. Speakers} & 16 & 11 & 21 & 13 & 24\\ \hline
\textbf{Avg. Speaker Dur.} & 23.04 s &  28.74 s & 25.18 s & 22.36 s & 17.12 s\\ \hline
\textbf{Speaker Dur. (Std. Dev)} & 6.72 s & 6.08 s & 5.58 s & 5.65 s & 4.97 s\\ \hline 
\end{tabular}}
\caption{Speaker Statistics for Dataset}
\label{tab:spkStat}
\end{table}

Tables \ref{tab:SNRvals} and \ref{tab:spkStat} are a general analysis of the 100 hours, aiming to shed some light on the properties of the data. Average number of speakers and the average and variation of speaker duration per 30 minute file of a channel is provided. In addition, the average and variation of SNR within each channel file is also displayed. Researchers will be provided with the channel information for the released data after the Challenge concludes.

\subsection{Development Set}
For all tasks with the exception of SID, the Dev set consists of a total duration of 20 hours and 10 minutes and consists of around 60\% audio from clean channels and the other 40\% is from degraded channels. For the SID task, a separate Dev set is provided. (See Section .\ref{subsec:SIDDevSet})
\subsection{Training Set}
Around 60 hours of audio data will be provided for which baseline system generated sentiment labels will be provided. Detailed information regarding the baseline systems and results will be released in a separate document. Researchers may use this data as they see fit.
\subsection{Evaluation Set}
\label{sec:Timeline}
Only the audio files will be provided for evaluation. The Eval set will consist of similar amounts of audio data of every channel, comprising of 20 hours in total. The helpful statistics about the Eval set are given in Tables \ref{tab:spkStat} and \ref{tab:SNRvals}.

\section{Challenge Tasks}
As an effort to motivate an initial research direction, five core speech and language tasks. These following Tasks are designed to advance research efforts not only in speech processing and machine learning, but also in natural language understanding.
These five tasks include selected speech segments which would have the highest impact in terms of speech, text and language analysis. As the focus of this Challenge is mostly core speech tasks, speech content from both primary and background conversations is considered to form ground truth labels.
\subsection{TASK\#1: Speech Activity Detection (SAD)}
The goal in the SAD task is to automatically detect the presence of speech segments in audio recordings of variable duration. A system output is scored by comparing the system produced start and end times of speech and non-speech segments in audio recordings to human annotated start and end times. Correct, incorrect, and partially correct segments will determine error probabilities for systems and will be used to measure a system’s SAD performance by calculating the Detection Cost Function (DCF) value. The DCF is a function of false-positive (false alarm) and false-negative (missed detection) rates calculated from comparison to the human annotation that will be the reference for the comparison\footnote{\url{https://www.nist.gov/itl/iad/mig/opensat}}. The goal for system developers will be to determine and select their system detection threshold, $\theta$, that minimizes the DCF value.

\subsection{TASK\#2: Speaker Diarization (SD)}

Speaker diarization has received much attention by the speech community, and while there are many available state-of-the-art systems for telephone speech, broadcast news and meetings, their performance does not translate to naturalistic speech in highly degraded noise environments. Some of the challenges diarization systems can encounter with the Apollo data are mentioned in Section \ref{subsec:DataChallenges}. This challenge is focused on Diarization from scratch.  

\subsection{TASK\#3: Speaker Identification (SID)}
\label{subsec:SIDDevSet}
In addition to the issues faced by diarization systems, Speaker Identification system performance also relies on speech content per segment. Contiguous speech by a single speaker between 0.4 and 50 seconds have been observed in this data, and a significant portion of short utterances exist in the Corpus. With over 350 known speakers contributing in varying degree of content, the sample set of speakers is narrowed down to 183 speakers with at least 10 seconds of total speech content, that are distributed in the Dev and Eval Sets. Table \ref{tb:SIDtab}. displays the necessary information.

\begin{table}[!ht]
\centering
{\renewcommand{\arraystretch}{1.2}%
\begin{tabular}{ |c|c|c|c|c|c| }
\hline
\textbf{Dataset} & \textbf{No. of speakers} & \textbf{ Avg. dur/spkr}  & \textbf{ Median dur/spkr}  & \textbf{ Avg. dur/utt} & \textbf{Total utts.}\\ \hline
Dev  & 183 & 247.7\,s & 50.38\,s & 5.35\,s & 8394\\ \hline
Eval & 183 & 105.3\,s & 21.42\,s & 4.69\,s & 3600\\ \hline
\end{tabular}}
\caption{SID Dev and Eval Sets}
\label{tb:SIDtab}
\end{table}

The primary focus of this challenge will be in-set identification of speakers with drastically varying duration of speech. A simple Top-5 accuracy metric to gauge system performance is mentioned in Section \ref{subsec:SID}. 

\subsection{TASK\#4: Automatic Speech Recognition (ASR)}
The goal of the ASR task is to automatically produce a verbatim, case-insensitive transcript of all words spoken in an audio recording. ASR performance is measured by the word error rate (WER), calculated as the sum of errors (deletions, insertions and substitutions) divided by the total number of words from the reference. Sections of audio which could not be labeled by manual annotators will be provided in a separate folder. These sections will not be considered for scoring. This segment information will be updated on the official Challenge website.

\subsection{TASK\#5: Sentiment Detection (SENTIMENT)}
Speech based sentiment extraction is an emerging and challenging field. In the context of conversational speech understanding, sentiment expressed by a speaker will help to learn the speaker’s opinion. Coupled with Topic detection to build a joint topic and sentiment detecting system will help to understand various aspects of conversational behavior of speakers. When compared to premeditated recorded content (for example, YouTube videos, Amazon reviews); normal day-to-day conversations tend to display sentiment in a very subtle way. This makes the problem very challenging. Trying to detect sentiment and use them to understand conversational behavior in the complex speech corpora like the Fearless Apollo corpus will help to push the limits of conversational AI.
Building sentiment detection system for Fearless Apollo data poses challenges like:
\begin{enumerate}
    \item Domain and vocabulary: The speaker can express opinions about any topic, (e.g., technical, products, social, games, day-to-day issues etc.) Hence the ASR system should be efficient to handle a wide range of domains and vocabulary. The language model should be comprehensive to take care of technical as well as mundane conversations.
    \item Speaker variability and speaker accents: ASR system should be robust to speaker variability which includes a wide range of English accents.
    \item Noisy audio and channels: Apollo data was digitized recently from 30 track analog tapes whose data was recoded around 50 years ago. Most of the audio channels of Fearless Apollo corpus suffer from different issues (one or many) like high channel noise, system noise, attenuated signal bandwidth, ransmission noise, cosmic noise, analog tape static noise, noise from tape aging, etc,. Some channels are noisier than others. More details in the audio channels and the data variability issues.
    \item Natural and Spontaneous: Detecting audio sentiment in natural and spontaneous speaker settings and various speaker interactive scenarios (i.e., 1-way, 2-way, public speech etc.) is challenging. In our scenario, The speakers are conversing between ground and space. There can be very tensed/exited moments where there is rapid conversational turns, with stressful or calm situations. Many a times the conversations can also be very short (may be, one word). These can pose challenge for both ASR, diarization and sentiment detection systems.
    
\end{enumerate}

\subsection{Performance Metrics}
We have followed NIST recommended standards for SAD, SD and ASR tasks, and have kept a simple Top-S accuracy measure for SID Task.

\vspace{1em}
\noindent \textbf{SAD} \par
Four system output possibilities are considered:
\begin{enumerate}
    \item True Positive (TP) -- system correctly identifies start-stop times of speech segments compared to the reference (manual annotation),
    \item True Negative (TN) -- system correctly identifies start-stop times of non-speech segments compared to reference,
    \item False Positive (FP) -- system incorrectly identifies speech in a segment where the reference identifies the segment as non-speech, and
    \item False Negative (FN) -- system missed identification of speech in a segment where the reference identifies a segment as speech.
\end{enumerate}

\begin{figure}[!ht]
\includegraphics[width=17cm, height=9.72cm]{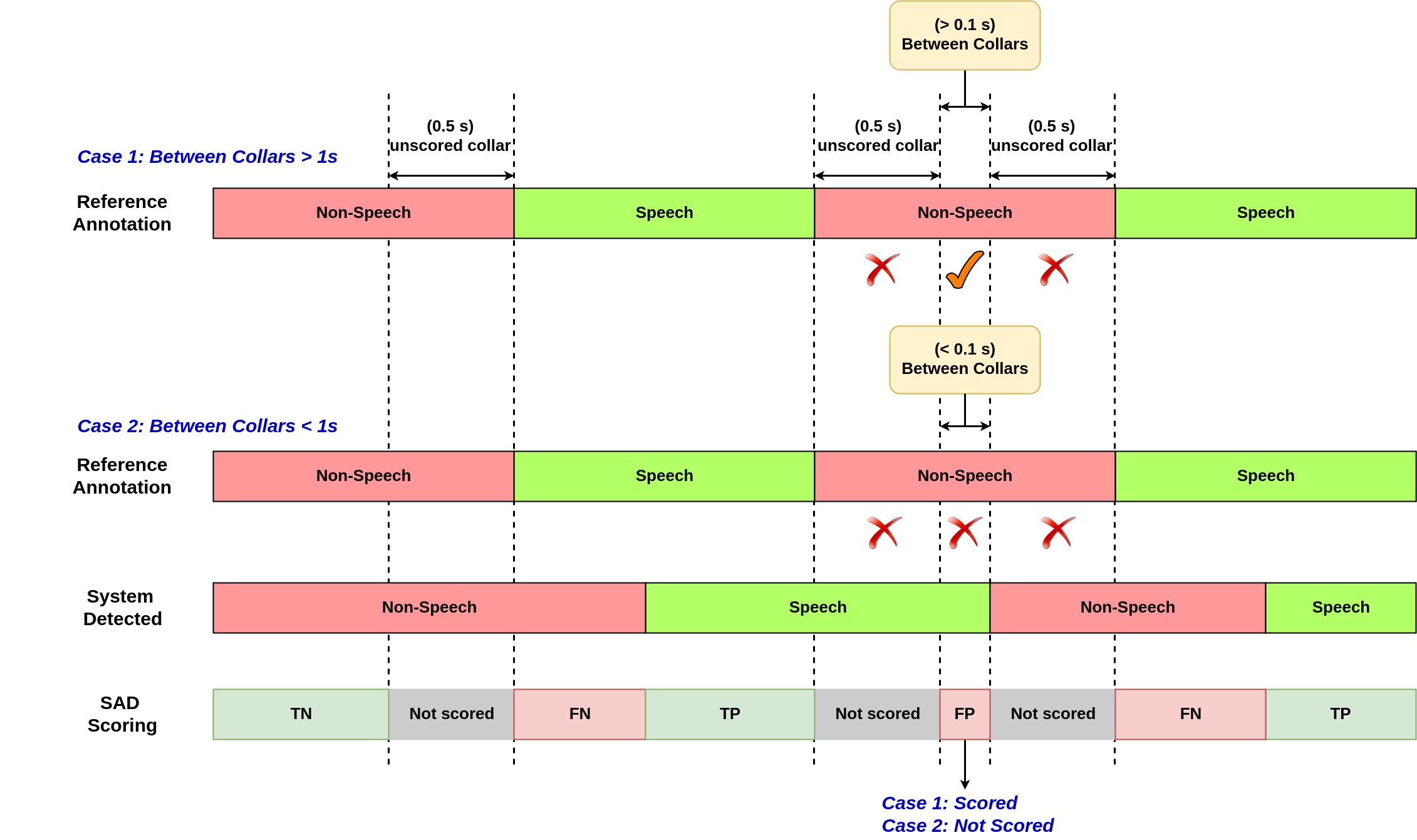}
\caption{SAD Scoring and Collar Information}
\label{fig:SAD_rules}
\end{figure}

SAD error rates represent a measure of the amount of time that is misclassified by the system’s segmentation of the test audio files. Missing, or failing to detect, actual speech is considered a more serious error than misidentifying its start and end times. A 0.5\,s collar, a “buffer zone”, at the beginning and end of each speech region will not be scored, taking into account inconsistencies in human annotations. If a segment of non-speech between collars is not 0.1\,s or greater, then the collars involved are expanded to include the less-than 0.1\,s non-speech. For example, no resulting non-speech segment with a duration of just 0.099\,s can exist. Similarly, for a region of non-speech before a collar at the beginning of the file or a region of non-speech after a collar at the end of the file, the resulting non-speech segment must last at least 0.1\,s or else the collar will expand to include it. In all other circumstances the collars will be exactly the nominal length. 
Figure \ref{fig:SAD_rules}: Shows how a \textless 0.1\,s non-speech segment (0.09 s) is added to the collars and not used in scoring, and illustrates the relationship between human annotation, the scoring regions resulting from application of the collars, a hypothetical system detected output, and the resulting time intervals from the four system output possibilities.

The scoring collars also help compensate for ambiguities in noisy channel annotation. Non-speech collars of half a second in length will define those regions that will not be scored. As can be seen, with collars applied to the annotation, parts of system-detected non-speech and potentially speech are not used in scoring. Below illustrates an example of a system detected output and the resulting scoring zones relative to the annotation with 0.5\,s collars applied. The figure shows the resulting four possibilities (TN, FN, TP, FP) considered in the scoring. The gray areas preceding and trailing the annotated speech are the 0.5\,s collar regions.


\vspace{1em}
\textbf{Scoring Procedure}: 
Information for downloading the scoring software will be available at the OpenSAT website. The four system output possibilities mentioned above determine the probability of a false positive (P$_{FP}$ ) and the probability of a false negative (P$_{FN}$ ). Developers are responsible for determining a hypothetical optimum setting (${\theta}$) for their system that minimizes the DCF value.

\vspace{2em}
P$_{FP}$ = detecting speech where there is no speech, also called a “false alarm”; \\
P$_{FN}$ = missed detection of speech, i.e., not detecting speech where there is speech, also called a “miss”;

\[P_{FP} = \frac{\textit{total FP time}}{\textit{annotated total nonspeech time}}\] 
\[P_{FN} = \frac{\textit{total FN time}}{\textit{annotated total speech time}}\] 

DCF($\theta$) is the detection cost function value for a system at a given system decision-threshold setting

\[DCF(\theta) = 0.75 \times P_{FN}(\theta) + 0.25 \times P_{FP}(\theta)\]

P$_{FN}$ and P$_{FP}$ are weighted by 0.75 and 0.25, respectively, $\theta$  denotes a given system decision-threshold setting.

\vspace{2em}
\par \noindent \textbf{SD}\label{subsec:SD} \par 

Diarization error rate (DER), introduced for the NIST Rich Transcription Spring 2003 Evaluation (RT-03S)\footnote{\url{https://catalog.ldc.upenn.edu/LDC2007S10}},
is the total percentage of reference speaker time that is not correctly attributed to a speaker, where “correctly attributed” is defined in terms of an optimal one-to-one mapping between the reference and system speakers.
More concretely, DER is defined as:

\[DER = \frac{(FA + MISS + ERROR)}{TOTAL}\] 

where
\begin{itemize}
    \item \textbf{TOTAL} is the total reference speaker time; that is, the sum of the durations of all reference speaker
    \item \textbf{FA} is the total system speaker time not attributed to a reference speaker
    \item \textbf{MISS} is the total reference speaker time not attributed to a system speaker
    \item \textbf{ERROR} is the total reference speaker time attributed to the wrong speaker segments 
\end{itemize}

\vspace{1em}
The recommended open source scoring tool is maintained as a github repo\footnote{\url{https://github.com/nryant/dscore}}. \newline
To score a set of system output RTTMs \textit{dev\_001.rttm} against corresponding reference RTTMs \textit{ref\_001.rttm} using the un-partitioned evaluation map (UEM) \textit{dev\_001.uem}, the command line would be: \\ \vspace{0.5em}

\begin{Verbatim}[fontsize=\large]
 python score.py -u dev_001.uem -r ref_dev_001.rttm  -s dev_001.rttm 
\end{Verbatim}

\vspace{1em}
The per-file results for DER will be considered for evaluation. For additional details about scoring tool usage, please consult the documentation given in the github repository. \\

\vspace{2em}
\textbf{Labeling Criterion:} \par
\begin{itemize}
    \item Speech segments for which human annotators were unable to identify the the Speaker have been labeled ``UNK" and will not be scored 
    \item Overlap instances will not be considered for Diarization 
    \item All the scoring regions will be provided in the UEM files, (refer to Section  \ref{appendix:UEM}). For the dev set, the UEM will be available for download.
    \item Consecutive segments of the same speaker with a non-speech segment of $\leq$ 1 second come together and are considered as a single segment. 
    \item A forgiveness collar of 0.25 seconds, before and after each reference boundary, will be considered in order to take into  account  both  inconsistent  human  annotations  and  the  uncertainty about when a speaker begins or ends.
\end{itemize}

\vspace{2em}
\par \noindent \textbf{SID}\label{subsec:SID} \par
The SID Task will be evaluated for Accuracy of the Top-5 system predictions for a given input file. 

\[Accuracy = \frac{\sum_{i \in S} N_{sys}(i)}{\sum_{i=1}^{M} N_{ref}(i)} ,\hspace{2em} S=\{k \in [1,M] : N_{ref}(k) \subseteq N_{sys}(k)\} \] 

where,\\ 
$N_{ref}(i)$ = speaker labels from ground truth for $i^{th}$ segment, \\
$N_{sys}(i)$ = system predicted speaker labels for $i^{th}$ segment, \\
$M$ = total number of segments \\

\textbf{Labeling Criterion:} \par
\begin{itemize}
    \item Speech segments for which human annotators were unable to identify the Speaker have been labeled ``UNK" and will not be provided as instances in Dev the Eval sets.
    \item Speech segments which have an associated speaker sabel may contain background noise and presence of speech from background conversations.
    \item Consecutive segments of the same speaker with a non-speech segment of $\leq$ 2 seconds come together and are considered as a single segment.
\end{itemize}

\vspace{3em}
\par \noindent \textbf{ASR} \par

Manual transcription for the Dev and Eval sets has been carried out using hand annotated SAD labels as a starting point.

\vspace{1em}
System performance metric computation
\begin{itemize}
    \item An overall Word Error Rate (WER) will be computed as the fraction of token recognition errors
per maximum number of reference tokens (scorable and optionally deletable tokens):
\end{itemize}

\[WER = \frac{(N_{Del} + N_{Ins} + N_{subst})}{N_{Ref}}\] 

where,

N$_{Del}$ = number of unmapped reference tokens (tokens missed, not detected, by the system) \\
N$_{Ins}$ = number of unmapped system outputs tokens (tokens that are not in the reference) \\
N$_{Subst}$ = number of system output tokens mapped to reference tokens but non-matching to the
reference spelling \\
N$_{Ref}$ = the maximum number of reference tokens (includes scorable and optionally deletable
reference tokens) \par
The tool for WER calculation is provided by Kaldi \footnote{\url{http://kaldi-asr.org/doc/tools.html}}
\vspace{1em}

\textbf{Labeling Criterion:} \par
\begin{itemize}
    \item Segments or parts of speech unintelligible to the annotators are marked as [unk] and will not be scored during evaluation
    \item Overlap instances with SNR $\leq$ 0 dB are marked as [unk] and will not be scored during evaluation
    \item Consecutive segments of the same speaker with a silent of less that 2 seconds come together and are considered as a single segment.
\end{itemize}

\vspace{2em}

\par \noindent \textbf{SENTIMENT} \par
\vspace{1em}
In this challenge, we would propose the teams to develop a sentiment detection system for naturalist speech. Example sentiment systems can be referred at [1][2].  The sentiment system developed should detect 3 sentiment outcome polarities, namely

\begin{itemize}
    \item POSITIVE
    \item NEUTRAL
    \item NEGATIVE
\end{itemize}
\vspace{1em}
For the Sentiment Task, 20 hours of development set audio, corresponding hand curated transcripts and ASR generated transcripts are provided, and another 20 hours of the evaluation set will be used for scoring systems.
The following Accuracy metric is used for scoring this Task.
\vspace{1em}
\[Acc_{sent} = \frac{\textit{total TP time}}{\textit{annotated total speech time}}\] 
\vspace{1em}

\textbf{Labeling Criterion:} \par
\begin{itemize}
    \item Every 10ms ground-truth speech segment will be scored and a cumulative accuracy score over the entire scoring region will be provided for each audio file.
    \item Non-Speech segments will not be scored.
    \item Speech segments for which human annotators were unable to identify the the speech content have been labeled as ``[unk]" and will not be scored .
    \item Overlap instances will not be scored
    \item The development set is provided as a JSON file per audio file, and the same format will be expected for system submissions on the evaluation set (refer to Section \ref{appendix:JSON}).
    \item Consecutive segments of the same speaker with a non-speech segment of $\leq$ 1 second come together and are considered as a single segment. 
    \item A forgiveness collar of 2 seconds, before and after each reference boundary, will be considered in order to take into  account  both  inconsistent  human  annotations  and  the  uncertainty about when a speaker begins or ends.
\end{itemize}

\section{Evaluation Conditions}
\label{subsec:DataRelease}

In the event that researchers have access to the entire Fearless Steps Corpus, they may not use the audio data corresponding to the following tapes for any task: 868, 869, 870, 883, 884, 885, 886\footnote{Tape Identification Numbers labeled by NASA}.
Apart from the aforementioned data, Researchers may use any other data of their choice for system training and development. 
All the challenge data to be released (audio and transcriptions) are given in Table. \ref{tab:TabdataRelease}.

\begin{table}[!ht]
\centering
{\renewcommand{\arraystretch}{1.2}%
\begin{tabular}{ |l|l| }
\hline
\textbf{Dataset Release} & \textbf{Tasks}\\ \hline
Data $\rightarrow$ Tracks $\rightarrow$ Train  & SID, ASR, SENTIMENT \\ \hline
Data $\rightarrow$ Tracks $\rightarrow$ Dev  & SAD, SD, ASR, SENTIMENT \\ \hline
Eval $\rightarrow$ Data $\rightarrow$ Tracks & SAD, SD, ASR, SENTIMENT \\ \hline
Data $\rightarrow$ Speakers $\rightarrow$ Train & SID \\ \hline
Eval $\rightarrow$ Data $\rightarrow$ Speakers & SID \\ \hline
\end{tabular}}
\caption{Task-Data assignment description}
\label{tab:TabdataRelease}
\end{table}

\section{Evaluation Rules}

\begin{enumerate}
    \item Site registration will be required in order to participate
    \item Researchers who register but do not submit a system to the Challenge are considered withdrawn from the Challenge
    \item Researchers may use any audio and transcriptions to build their systems, with the exception of the data mentioned in Section \ref{subsec:DataRelease}.
    \item Only the audio for the blind eval set (20 hours) will be released. Researchers are expected to run their systems on the blind eval set. 
    \item Investigation of the evaluation data prior to submission of all systems outputs is not allowed. Human probing is prohibited.
\end{enumerate}
While participants are encouraged to submit papers to the special session at Interspeech 2019, this is not a
requirement for participation.

\section{Evaluation Protocol}
\begin{itemize}
    \item  The entire Fearless Steps Corpus (consisting of over 11,000 hours of audio from the Apollo-11 Mission) including the 100 hours is publicly available and requires no additional licence to use the data.
    \item There is no cost to participate in the Fearless Steps evaluation. Development data and evaluation data will be freely made available to registered participants.
    \item At least one participant from each team must register on the \href{http://fearlesssteps.exploreapollo.org/}{Fearless Steps Challenge 2019}.
    \item System output submissions will be sent to the official \href{FearlessSteps@utdallas.edu}{Fearless Steps} correspondence email-id. See Appendix. \ref{appen:SysDesc} for system output submission packaging.
    \item Participants can submit at most 2 system submissions per day.
    \item Results of submitted systems will be mailed to the registered email-id within a week of the submission. 
    \item It is required that participants agree to process the data in accordance with the following rules.
\end{itemize}

\section{System Input and Output}
\subsection{Audio Files}
All audio files provided will be in the standard single channel (mono) 16-bit PCM `WAV' format, sampled at 8000 Hz. \par
\noindent
\textbf{For SAD, SD, ASR and SENTIMENT Tasks}: Each file in the Train, Dev, and Eval set have a duration between 30 and 33 minutes. There are no other exceptions. \par
\noindent
\textbf{For SID Task}: The Dev and Eval sets for this task have segmented Files with a single speaker information. The associated speaker label for the Dev set will be provided in the file name. These files may have a duration ranging from 2.2 to 15 seconds (inclusive).
\subsection{Speech Activity Detection (SAD)}
For both Input and Output file format descriptions, see Appendix \href{appendix:SADFF}{A.3}

\subsection{Speaker Diarization (SD)}
For the non-scoring region Input File format description, see Appendix \href{appendix:UEM}{B.1} \par
For both Input and Output file format descriptions, see Appendix \href{appendix:RTTM}{A.1}
\subsection{Speaker Identification (SID)}
For Output file format description, see Appendix \href{appendix:SIDFF}{A.4}
\subsection{Automatic Speech Recognition (ASR)}
For both Input and Output file format descriptions, see Appendix \href{appendix:JSON}{A.2}
\subsection{Sentiment Detection (SENTIMENT)}
For both Input and Output file format descriptions, see Appendix \href{appendix:JSON}{A.2}

\section{Updates}
\vspace{-1em}
\justifying
During registration, researchers are advised to provide an email address they are most active on. Any other updates and changes will be displayed on the website (\href{http://fearlesssteps.exploreapollo.org/}{Fearless Steps Challenge}) and sent through a mailing list to all registered participants. Contact \href{FearlessSteps@utdallas.edu}{Fearless Steps} for questions relevant to FS-1 2019 not covered in this evaluation plan.

\section{Interspeech 2019 Special Session}
\vspace{-1em}
\justifying
The results of the challenge will be presented at a special session at Interspeech 2019, held from September 15$^{th}$ - 19$^{th}$, 2019 in Graz, Austria. Researchers wishing to submit papers should do so through the Interspeech submission portal. Additional instructions will be provided once the Interspeech submission portal opens.
The Challenge will continue as planned through 28$^{th}$ June 2019. Researchers will be notified privately of their final system ranking, on or before 1$^{th}$  July (deadline for submission of the camera ready paper). System rankings will be announced publicly on the 50$^{th}$ Anniversary of the first moon walk. 

\section{Acknowledgements}
\vspace{-1em}
\justifying
This project was funded by AFRL under contract FA8750-15-1-0205, NSF Project 1219130, and partially by the University of Texas at Dallas from the Distinguished University Chair in Telecommunications Engineering held by J. H. L. Hansen. UTD-CRSS lab would like to thank Gregory H. Wiseman and Karen L. Walsemann from National Aeronautics and Space Administration (NASA) and Tuan Nguyen for their never ending support throughout the digitization process. 

\clearpage
\begin{appendices}
\normalsize
\section{Label File Format Specification}

\subsection{RTTM File Format}
\label{appendix:RTTM}
Systems should output their diarizations as Rich Transcription Time Marked (RTTM) files \footnote{\href{https://web.archive.org/web/20170119114252/http://www.itl.nist.gov/iad/mig/tests/rt/2009/docs/rt09-meeting-eval-plan-v2.pdf}{https://web.archive.org/web/20170119114252/http://www.itl.nist.gov/iad/mig/tests/rt/2009/docs/rt09-meeting-eval-plan-v2.pdf}}. A NIST defined File Format, the RTTM files are text files containing one turn per line, each line containing nine space-delimited fields:

\begin{description}[style=multiline,leftmargin=4.5cm,font=\normalfont]
    \item[\hspace{0.5cm}$\bullet$ \textbf{Type}]segment type; should always by ``SPEAKER"
    \item[\hspace{0.5cm}$\bullet$ \textbf{File ID}] file name; basename of the recording minus extension (e.g., ``FS\_P01\_eval\_023")
    \item[\hspace{0.5cm}$\bullet$ \textbf{Channel ID}] channel (1-indexed) that turn is on; should always be ``1"
    \item[\hspace{0.5cm}$\bullet$ \textbf{Turn Onset}] onset of turn in seconds from beginning of recording
    \item[\hspace{0.5cm}$\bullet$ \textbf{Turn Duration}] duration of turn in seconds
    \item[\hspace{0.5cm}$\bullet$ \textbf{Orthography Field}] should always by ``\textless NA\textgreater"
    \item[\hspace{0.5cm}$\bullet$ \textbf{Speaker Type}] should always be ``\textless NA\textgreater"
    \item[\hspace{0.5cm}$\bullet$ \textbf{Speaker Name}] name of speaker of turn; should be unique within scope of each file
    \item[\hspace{0.5cm}$\bullet$ \textbf{\textit{Confidence Score}} ]  (\textit{Optional}) system confidence (probability) that information is correct; should always be {\textless NA\textgreater}
\end{description}

\noindent
\textbf{For instance:} \par \par
\indent
\texttt{SPEAKER FS\_P01\_dev\_001 1 256.04 2.35 {\textless NA\textgreater} {\textless NA\textgreater}  EECOM1 {\textless NA\textgreater}}

\texttt{SPEAKER FS\_P01\_dev\_001 1 358.08 3.06 {\textless NA\textgreater} {\textless NA\textgreater} FD1 {\textless NA\textgreater}}

\texttt{SPEAKER FS\_P01\_dev\_001 1 368.97 2.22 {\textless NA\textgreater} {\textless NA\textgreater} GNC1 {\textless NA\textgreater}}

\clearpage
\subsection{JSON File Format}
\label{appendix:JSON}
The transcriptions are provided in JSON format for each file as \textless file\_ID\textgreater.json \footnote{\href{https://www.json.org/}{https://www.json.org/}}. The JSON file includes the following pieces of information for each utterance:

\begin{description}[style=multiline,leftmargin=4cm,font=\normalfont]
    \item[\hspace{0.5cm}$\bullet$ \textbf{Speaker ID}] Token: ``speakerID"
    \item[\hspace{0.5cm}$\bullet$ \textbf{Transcription}] Token: ``words"
    \item[\hspace{0.5cm}$\bullet$ \textbf{Sentiment}] Token: ``sentiment"
    \item[\hspace{0.5cm}$\bullet$ \textbf{Start Time}] Token: ``startTime"
    \item[\hspace{0.5cm}$\bullet$ \textbf{End Time}] Token: ``endTime"
\end{description}

\noindent
\textbf{For instance:} \par \par
\indent
\texttt{\{ 
\begin{description}[style=multiline,leftmargin=4cm,font=\normalfont]
    \item[\hspace{1cm} \texttt{"speakerID":}] "NEIL", 
    \item[\hspace{1cm} \texttt{"words":}] "It's one small step for man,",
    \item[\hspace{1cm} \texttt{"sentiment":}] "POSITIVE" 
    \item[\hspace{1cm} \texttt{"startTIme":}] "1325.203"
    \item[\hspace{1cm} \texttt{"endTime":}] "1327.501"
\end{description}
\},\newline
\{ 
\begin{description}[style=multiline,leftmargin=4cm,font=\normalfont]
    \item[\hspace{1cm} \texttt{"speakerID":}] "NEIL", 
    \item[\hspace{1cm} \texttt{"words":}] "One, Giant leap for mankind.",
    \item[\hspace{1cm} \texttt{"sentiment":}] "POSITIVE" 
    \item[\hspace{1cm} \texttt{"startTIme":}]  "1330.162"
    \item[\hspace{1cm} \texttt{"endTime":}] "1332.89" 
\end{description}
\}, ....}

\clearpage
\subsection{SAD File Format}
\label{appendix:SADFF}
Systems should output their SAD as text (txt) files. A NIST defined File Format, the text files are text files containing one turn per line, each line containing nine tab-delimited fields:

\begin{description}[style=multiline,leftmargin=4cm,font=\normalfont]
    \item[\hspace{0.5cm}$\bullet$ \textbf{Test}] Test Definition File Name (Value: X)
    \item[\hspace{0.5cm}$\bullet$ \textbf{TestSet ID}] contents of the id attribute TestSet tag (Value: X)
    \item[\hspace{0.5cm}$\bullet$ \textbf{Test ID}] contents of the id attribute of the TEST tag (Value: X)
    \item[\hspace{0.5cm}$\bullet$ \textbf{Task}] SAD $<$== a literal text string, without quotations (Value: SAD)
    \item[\hspace{0.5cm}$\bullet$ \textbf{File ID}] contents of the id attribute of the File tag (Value: X)
    \item[\hspace{0.5cm}$\bullet$ \textbf{Interval start}] an offset, in seconds from the start of the audio file for the \textbf{start} of the speech/non-speech interval (Value: floating number)
    \item[\hspace{0.5cm}$\bullet$ \textbf{Interval end}]  an offset, in seconds from the start of the audio file for the \textbf{end} of the speech/non-speech interval (Value: floating number)
    \item[\hspace{0.5cm}$\bullet$ \textbf{Type}] In system output: speech/non-speech without quotation marks (Value: speech/non-speech)\\In the reference: S/NS for speech/non-speech
    \item[\hspace{0.5cm}$\bullet$ \textbf{\textit{Confidence Score}} ] (\textit{Optional}) A value in the range 0 thorugh 1.0, with higher values indicating greater confidence about the presence/absence of speech 
\end{description}

\noindent
\textbf{For instance:} SAD system output file
\begin{table}[!ht]
\texttt{
\begin{tabular}{lllllllll}
X & X & X & SAD &  X & 0.00 & 5.77 & speech & 0.500000 \\
X & X & X & SAD & X & 5.77 & 6.37 & non-speech & 0.500000 \\
X & X & X & SAD & X & 6.37 & 11.22 & speech & 0.500000\\
\end{tabular}
}
\end{table}

Interval overlapping will be disallowed and will fail in validating your files. Example of overlapping:

\begin{table}[!ht]
\texttt{
\begin{tabular}{lllllllll}
X & X & X & SAD &  X & 0.00 & \textbf{5.77} & speech & 0.500000 \\
X & X & X & SAD & X & \textbf{5.13} & 6.37 & non-speech & 0.500000 \\
\end{tabular}
}
\end{table}

\clearpage
\subsection{SID Output File Format}
\label{appendix:SIDFF}
The SID output file should be a text file containing one test-segment per line, each line containing five space-delimited fields

\begin{description}[style=multiline,leftmargin=4cm,font=\normalfont]
    \item[\hspace{0.5cm}$\bullet$ \textbf{Test}] Test Definition File Name
    \item[\hspace{0.5cm}$\bullet$ \textbf{Prediction 1}] Top System SpeakerID Prediction
    \item[\hspace{0.5cm}$\bullet$ \textbf{Prediction 2}] 2nd Most Likely System SpeakerID Prediction
    \item[\hspace{0.5cm}$\bullet$ \textbf{Prediction 3}] 3rd Most Likely System SpeakerID Prediction
    \item[\hspace{0.5cm}$\bullet$ \textbf{Prediction 4}] 4th Most Likely System SpeakerID Prediction
    \item[\hspace{0.5cm}$\bullet$ \textbf{Prediction 5}] 5th Most Likely System SpeakerID Prediction
\end{description}

\noindent
\textbf{For instance:} SID system output file
\begin{table}[!ht]
\texttt{\
\begin{tabular}{llllll}
FS\_P01\_dev\_FD1\_001 & FD1 & GNC1 & INCO &  NEIL & BUZZ \\
FS\_P01\_dev\_NEIL\_025 & FD1 & AGC & FIDO2 &  NEIL & BUZZ \\
FS\_P01\_eval\_001 & FD1 & GNC2 & INCO &  NEIL & BUZZ \\
FS\_P01\_eval\_005 & FIDO & GNC1 & GUIDANCE &  NEIL & BUZZ \\
\end{tabular}
}
\end{table}

\textbf{File Naming conventions:}
No additional text file is provided for the speaker identification labels. Every segment consists of one speaker to evaluate. Researchers are expected to retrieve the Speaker label for each segment from the file name. \\ \\

\textbf{Development Set:}\\ \\
\texttt{FS\_P01\_dev\_\textless Speaker ID\textgreater\_\textless Dev Utterance ID\textgreater}\\

\vspace{2em}
\textbf{Evaluation Set:}\\ \\
\texttt{FS\_P01\_eval\_\textless Eval Utterance ID\textgreater}

\clearpage
\section{Supplementary File Format Specification}
The Supplementary Files are any files in addition to the audio and (system and ground-truth) label files that are necessary to evaluate the performance of a system accurately for all given Tasks.
\subsection{UEM File Format}
\label{appendix:UEM}
\normalsize
The scoring region for each audio file is provided separately through the NIST defined UEM format\footnote{\href{https://web.archive.org/web/20170119114252/http://www.itl.nist.gov/iad/mig/tests/rt/2009/docs/rt09-meeting-eval-plan-v2.pdf}{https://web.archive.org/web/20170119114252/http://www.itl.nist.gov/iad/mig/tests/rt/2009/docs/rt09-meeting-eval-plan-v2.pdf}}. These regions will be provided to the scoring tool via UEM files.
Un-partitioned evaluation map (UEM)\footnote{\href{https://catalog.ldc.upenn.edu/docs/LDC2004S11/readme.html\#INPUT_UEM}{https://catalog.ldc.upenn.edu/docs/LDC2004S11/readme.html\#INPUT\_UEM}} files are used to specify the scoring regions within each recording. For each scoring region, the UEM file contains a line with the following four space-delimited fields

\begin{description}[style=multiline,leftmargin=4cm,font=\normalfont]
    \item[\hspace{0.5cm}$\bullet$ \textbf{File ID}] file name; basename of the recording minus extension (e.g., ``FS\_P01\_dev\_001") 
    \item[\hspace{0.5cm}$\bullet$ \textbf{Channel ID}] channel (1-indexed) that scoring region is on 
    \item[\hspace{0.5cm}$\bullet$ \textbf{Onset}]  onset of scoring region in seconds from beginning of recording
    \item[\hspace{0.5cm}$\bullet$ \textbf{Offset}] offset of scoring region in seconds from beginning of recording
\end{description}

\noindent
\textbf{For instance:} \par \par
\begin{table}[!ht]
\texttt{\
\begin{tabular}{llll}
FS\_P01\_dev\_001 & 1 & 234.32 & 737.54  \\
FS\_P01\_eval\_001 & 1 & 832.35 &  1800.00 \\
FS\_P01\_eval\_007 & 1 & 124.46 &  624.23\\
\end{tabular}
}
\end{table}

\clearpage
\section{Data Resources for Training}
\normalsize

\vspace{1em}

This appendix identifies a (non-exhaustive) list of publicly available corpora the researchers may use for system training.\footnote{\href{https://www.ldc.upenn.edu/}{https://www.ldc.upenn.edu/}} \footnote{\href{https://coml.lscp.ens.fr/dihard/2018/data.html}{https://coml.lscp.ens.fr/dihard/2018/data.html}} 

\vspace{2em}

\noindent
\textbf{Corpora containing meeting speech} \par \par
\begin{itemize}
\item ICSI Meeting Speech Speech (LDC2004S02)
\item ICSI Meeting Transcripts (LDC2004T04)
\item ISL Meeting Speech Part 1 (LDC2004S05)
\item ISL Meeting Transcripts Part 1 (LDC2004T10)
\item NIST Meeting Pilot Corpus Speech (LDC2004S09)
\item NIST Meeting Pilot Corpus Transcripts and Metadata (LDC2004T13)
\item 2004 Spring NIST Rich Transcription (RT-04S) Development Data (LDC2007S11)
\item 2004 Spring NIST Rich Transcription (RT-04S) Evaluation Data (LDC2007S12)
\item 2006 NIST Spoken Term Detection Development Set (LDC2011S02)
\item 2006 NIST Spoken Term Detection Evaluation Set (LDC2011S03)
\item 2005 Spring NIST Rich Transcription (RT-05S) Evaluation Set (LDC2011S06)
\item Augmented Multiparty Interaction (AMI) Meeting Corpus (\href{http://groups.inf.ed.ac.uk/ami/corpus/}{http://groups.inf.ed.ac.uk/ami/corpus/})
\end{itemize}

\vspace{2em}

\noindent
\textbf{Conversational telephone speech (CTS) corpora} \par \par
\begin{itemize}
    \item Switchboard-1 Release 2 (LDC97S62)
    \item Fisher English Training Speech Part 1 Speech (LDC2004S13)
    \item Fisher English Training Speech Part 1 Transcripts (LDC2004T19)
    \item Arabic CTS Levantine Fisher Training Data Set 3, Speech (LDC2005S07)
    \item Fisher English Training Part 2, Speech (LDC2005S13)
    \item Arabic CTS Levantine Fisher Training Data Set 3, Transcripts (LDC2005T03)
    \item Fisher English Training Part 2, Transcripts (LDC2005T19)
    \item Fisher Levantine Arabic Conversational Telephone Speech (LDC2007S02)
    \item Fisher Levantine Arabic Conversational Telephone Speech, Transcripts (LDC2007T04)
    \item Fisher Spanish Speech (LDC2010S01)
    \item Fisher Spanish - Transcripts (LDC2010T04)
\end{itemize}

\vspace{2em}

\noindent
\textbf{Other corpora} \par \par
\begin{itemize}
    \item Speech in Noisy Environments (SPINE) Training Audio (LDC2000S87)
    \item Speech in Noisy Environments (SPINE) Evaluation Audio (LDC2000S96)
    \item Speech in Noisy Environments (SPINE) Training Transcripts (LDC2000T49)
    \item Speech in Noisy Environments (SPINE) Evaluation Transcripts (LDC2000T54)
    \item Speech in Noisy Environments (SPINE2) Part 1 Audio (LDC2001S04)
    \item Speech in Noisy Environments (SPINE2) Part 2 Audio (LDC2001S06)
    \item Speech in Noisy Environments (SPINE2) Part 3 Audio (LDC2001S08)
    \item Speech in Noisy Environments (SPINE2) Part 1 Transcripts (LDC2001T05)
    \item Speech in Noisy Environments (SPINE2) Part 2 Transcripts (LDC2001T07)
    \item Speech in Noisy Environments (SPINE2) Part 3 Transcripts (LDC2001T09)
    \item Santa Barbara Corpus of Spoken American English Part I (LDC2000S85)
    \item Santa Barbara Corpus of Spoken American English Part II (LDC2003S06)
    \item Santa Barbara Corpus of Spoken American English Part III (LDC2004S10)
    \item Santa Barbara Corpus of Spoken American English Part IV (LDC2005S25)
    \item HAVIC Pilot Transcription (LDC2016V01)
    \item LibriSpeech (\href{http://www.openslr.org/12/}{http://www.openslr.org/12/})
    \item VoxCeleb (\href{http://www.robots.ox.ac.uk/~vgg/data/voxceleb/}{http://www.robots.ox.ac.uk/vgg/data/voxceleb/})
\end{itemize}

\clearpage
\section{System descriptions}
\label{appen:SysDesc}
\normalsize
Proper interpretation of the evaluation results requires thorough documentation of each system. Consequently,at the end of the evaluation researchers must submit a full description of their system with sufficient detail for a fellow researcher to understand the approach and data/computational requirements. An acceptable
system description should include the following information:
\begin{itemize}
    \item Task
    \item Abstract
    \item Data resources
    \item Detailed description of algorithm
    \item Hardware requirements
\end{itemize}

\subsection*{Section 1: Task}
The Challenge Task for which the system is to be evaluated. Researchers can submit multiple systems for one or all of the tasks. However, each Submission should include a single system evaluation for a single task. If the same team of researchers wishes to submit multiple systems for a separate task, that submission should be done separately.

\subsection*{Section 2: Abstract}
A high-level description of the system.  

\subsection*{Section 3: Data resources}
This section should describe the data used for training including both volumes and sources.  For other publicly available corpora a link should be provided. In cases where a non-publicly available corpus is used, it should be described in sufficient detail
to get the gist of its composition. If the system is composed of multiple components and different components
are trained using different resources, there should be an accompanying description of which resources were
used for which components.

\subsection*{Section 4: Detailed description of algorithm}
Each component of the system should be described in sufficient detail that another researcher would be able to re-implement it. If hyperparameter tuning was performed,
there should be detailed description both of the tuning process and the final hyperparameters arrived at. We suggest including subsections for each major phase in the system. 

\clearpage
\subsection*{Section 4: Hardware requirements}
System developers should report the hardware requirements for both training and at test time.

\begin{itemize}
    \item Total number of CPU cores used
    \item Description of CPUs used (model, speed, number of cores)
    \item Total number of GPUs used
    \item Description of used GPUs (model, single precision TFLOPS, memory)
    \item Total available RAM
    \item Used disk storage
    \item Machine learning frameworks used (e.g., PyTorch, Tensorflow, CNTK)
\end{itemize}

System execution times to process a single 30 minute File must be reported. \\
For SID: system execution time for entire Eval Set must be reported
\end{appendices}
\clearpage


\end{document}